\documentclass[aps,prl,twocolumn,nopacs,superscriptaddress]{revtex4}
\usepackage{graphicx}
\usepackage{verbatim}
\usepackage{amsmath}
\usepackage{sidecap}
\usepackage{epstopdf}
\usepackage{braket}

\usepackage[usenames,dvipsnames]{xcolor}
\definecolor{Highlight}{rgb}{1,1,0.5}
\usepackage{soul}
\begin{document}

%\title{Evaluating the effects of mixed valence on resonant scattering in the vacuum ultraviolet to soft X-ray regimes}
\title{Second derivative analysis and alternative data filters for multi-dimensional spectroscopies: a Fourier-space perspective}
%\title{Improving on the Fourier-space profile of second derivative analysis}

%***add:
%Ting
%Jerzy T. Sadowski
%Takashi Taniguchi, Kenji Watanabe
%Advanced Materials Laboratory, National Institute of Materials Science, 1-1 Namiki Tsukuba, Ibaraki 305-0044, Japan

\author{Rongjie Li}
\author{Xiaoni Zhang}
\affiliation{Department of Physics, New York University, New York, New York 10003, USA}
\author{Lin Miao}
\affiliation{Department of Physics, New York University, New York, New York 10003, USA}
\affiliation{Advanced Light Source, Lawrence Berkeley National Laboratory, Berkeley, CA 94720, USA}
\author{Luca Stewart}
\author{Erica Kotta}
%\author{Yishuai Xu}
\affiliation{Department of Physics, New York University, New York, New York 10003, USA}
\author{Dong Qian}
\affiliation{Key Laboratory of Artificial Structures and Quantum Control (Ministry of Education), School of Physics and Astronomy, Shanghai Jiao Tong University, Shanghai 200240, China}
\affiliation{Collaborative Innovation Center of Advanced Microstructures, Nanjing 210093, China}
\author{Konstantine Kaznatcheev}
\author{Jerzy T. Sadowski}
\author{Elio Vescovo}
\affiliation{National Synchrotron Light Source II, Brookhaven National Laboratory, Upton, New York 11973, USA}
\author{Abdullah Alharbi}
\author{Ting Wu}
\affiliation{Electrical and Computer Engineering, New York University, Brooklyn, NY 11201}
\author{Takashi Taniguchi}
\author{Kenji Watanabe}
\affiliation{National Institute of Materials Science, 1-1 Namiki Tsukuba, Ibaraki 305-0044, Japan}
\author{Davood Shahrjerdi}
\affiliation{Electrical and Computer Engineering, New York University, Brooklyn, NY 11201}
\affiliation{Department of Physics, New York University, New York, New York 10003, USA}
%\author{Yishuai Xu}
\author{L. Andrew Wray}
\email{lawray@nyu.edu}
\thanks{Corresponding author}
\affiliation{Department of Physics, New York University, New York, New York 10003, USA}
%\affiliation{NYU-ECNU Institute of Physics at NYU Shanghai, 3663 Zhongshan Road North, Shanghai, 200062, China}

\begin{abstract}

The second derivative image (SDI) method is widely applied to sharpen dispersive data features in multi-dimensional spectroscopies such as angle resolved photoemission spectroscopy (ARPES). Here, the SDI function is represented in Fourier space, where it has the form of a multi-band pass filter. The interplay of the SDI procedure with undesirable noise and background features in ARPES data sets is reviewed, and it is shown that final image quality can be improved by eliminating higher Fourier harmonics of the SDI filter. We then discuss extensions of SDI-like band pass filters to higher dimensional data sets, and how one can create even more effective filters with some a priori knowledge of the spectral features.

\end{abstract}

% \pacs{73.20.At, 73.20.Hb, 72.10.Fk}

\date{\today}

\maketitle

\section{I. Introduction}

The past few decades have seen rapid improvement of throughput, resolution, and other fundamental capabilities for momentum-resolving spectroscopies such as angle resolved photoemission (ARPES) and resonant inelastic X-ray scattering (RIXS) \cite{Shen_Lu_ARPESreview, highHarmonic, Trions, AmentRIXSReview, WrayFrontiers}. The features observed with these techniques disperse in a multi-dimensional space with axes of momentum and energy, and can reveal the kinetics of quasiparticles that define the low energy properties of many-body systems. Moreover, the ongoing advancement of light source technologies has has driven technological developments that further increase the dimensionality of these spectroscopies by enabling space-, time-, spin-, and/or polarization- resolution \cite{Technology1,Technology2,Technology3,Technology4,Technology5,Technology6,Technology7,Technology8}. The second derivative image (SDI) analysis method is often applied to suppress undesirable backgrounds and sharpen the appearance of spectral features, and it has recently been noted that alternative analysis methods that take greater advantage of the full dimensionality of a data set can produce superior results in many cases \cite{ShenGradient,ZhangCurvature}. Here, we discuss the SDI method by representing it as Fourier space filter, and show that closely related one- and higher-dimensional band-pass filters can provide greater benefits for spectral analysis.

%These developments are particularly important in the context of the recent trends such as the investigation of very small or thin samples 

%use of low photon flux, and time/spin-resolved ARPES

%recent technical developments 

%applying band pass filtering technique will generally benefit ARPES under different conditions, i.e., high background cases, very small sample, low photon flux as well as related techniques such as time/spin resolved ARPES with relative low signal to noise ratio.

The need for SDI (and related) filters arrises when band features are difficult to observe on top of the incoherent background from multiple-scattering events \cite{Shirley_inelastic, BGgeneral, BGelastic}, or from other measurement limitations such as when studying a sample domain smaller than the incident photon beam. They are also used to sharpen features and make it easier to trace dispersions by eye. This paper focuses on two ARPES images as examples of these use-cases. The high-background case is an ARPES spectromicroscopy measurement performed with a $\lesssim20$ $\mu$m diameter beam spot on single-layer graphene transferred to hexagonal boronitride (hBN). This sample incorporates a suite of challenges that are increasingly common for ARPES investigations, including small `device-scale' dimensions, in-plane structural distortions from the layer transfer procedure, and a thickness that is smaller than the $\sim1$ nm \cite{UniversalCurve} ARPES penetration depth (weakening signal-to-background).

Furthermore, disorder was added to the graphene sample via low energy Ar ion bombardment, and the ARPES measurement was performed with a high photon energy (120 eV) that is relatively common to employ at spectromicroscopy beamlines, but not ideal for the study of graphene. Details of the sample preparation are described in Ref. \cite{DavoodAPL}. The measurement was performed at room temperature at the Brookhaven National Laboratory ESM beamline. The measurement resolution was relaxed to $\delta E \lesssim 30$ meV to enhance count rate, as the observed features were quite broad ($\gtrsim 500$ meV peak width at half maximum).

The other example is a measurement of multiple bands approaching the Fermi surface of IrTe$_2$. In this case, the measurement was performed on a large single crystal, and is signal-dominated. However the bands are not very well separated, and the SDI is used to obtain a cleaner image of their dispersions. Measurements on IrTe$_2$ were performed at the Advanced Light Source MERLIN beamline (BL 4.0.3), with 100 eV photons and energy resolution better than $\delta E \lesssim 20$ meV.

\begin{figure}[t]
\includegraphics[width = 8.7cm]{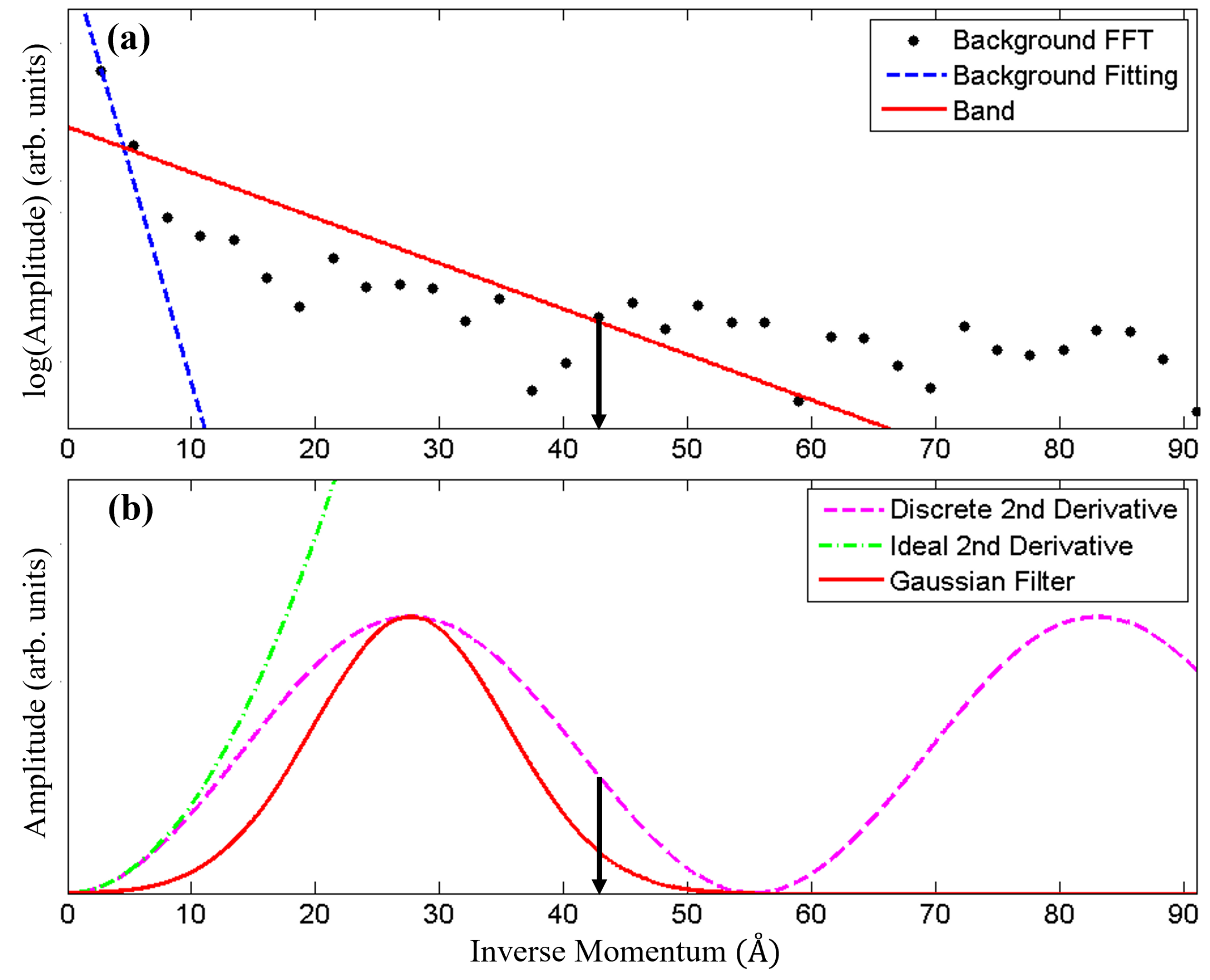}
\caption{{\bf{Data, noise and background in 1D Fourier space}}: (a) A semi-log graph showing (dots) the undesired background signal in an ARPES spectromicroscopy measurement on graphene (see raw data in Fig. 2(a)). The incoherent background is fitted as a blue dashed line, and the characteristic intensity profile of a real band feature is shown as a red solid line. A black arrow on the horizontal axis indicates the point at which the band feature falls beneath the noise floor. (b) The pass amplitude is shown for several Fourier space filters, including (green) an analytic second derivative, (magenta) a discrete numerical second derivative, and (red) a Gaussian filter. The second derivative and Gaussian filters are chosen to give a peak in the high signal-to-noise window from 5-40 \AA, while favoring higher frequencies that will yield a sharper image.}
\label{fig:schematics}
\end{figure} 

%Question: What's the feature width in Fig. 1.

\begin{figure*}[t]
\includegraphics[width = 17.5cm]{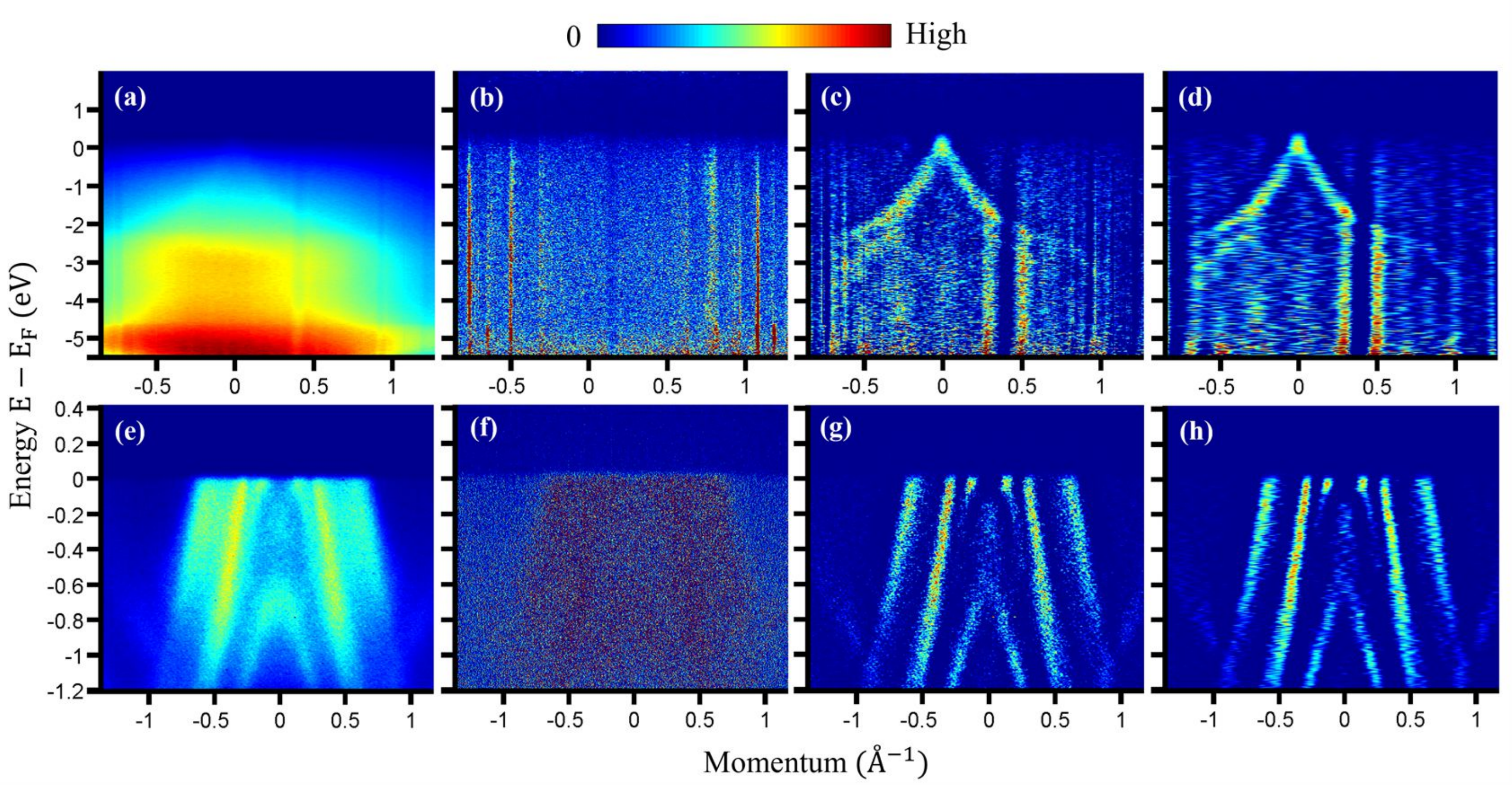}
\caption{{\bf{One-dimensional filters}}: (a) Raw ARPES spectromicroscopy data from a disordered graphene sample on hBN, with a very strong incoherent background. Momentum is zeroed on the K-point. (b-d) The data from panel (a) is filtered with (b) an analytic second derivative ($T(\xi)=\xi^2$), (c) the discrete numerical SDI method, and (d) a similar Gaussian filter (see Eq. \ref{Eq:corrGaussian} that eliminates higher Fourier harmonics of the discrete SDI. For convenience, the Gaussian uses the same centroid (c=27.9 \AA) as the first harmonic of the discrete SDI and a roughly identical width full width at half maximum (27 \AA). (e-h) The same analysis is presented for multiple band features seen near the $\Gamma$-point of IrTe$_2$.}
\label{fig:1Dfilters}
\end{figure*} 

%Edited to here
\section{II. ARPES data and filters in 1D Fourier space}

\subsection{II.A Signal, background and noise in Fourier space}

Figure 1(a) shows the Fourier decomposition of signal and background in a constant-energy slice (termed a momentum distribution curve, MDC) of ARPES data near the Fermi level of graphene, intersecting one of the Dirac cones.  From this spectrum, we can see that the background dominates the low frequency and high frequency sectors of Fourier space, while the Dirac band (red curve) has greater intensity in an intermediate window, from roughly 5-40 \AA. These trends are rather universal, and merely represent the fact that background features tend to be broader than real spectral features, giving them greater relative intensity at small frequencies, and statistical noise creates a flat `noise floor' in Fourier space that will dominate the spectrum at high frequencies. The universal nature of these features makes it easy to see that \emph{a band pass filter suppressing high and low frequencies} is likely to give a cleaner representation of data features seen by ARPES. We will show below that second derivative analysis already behaves as a form of band pass filter, and can be refined to function even better in this respect.

The decomposition of ARPES data into signal, background, and noise as presented in Fig. 1(a) is not obtainable from a direct measurement, and has been pieced together via several approximations.  The background distribution is measured from above the Fermi level, and has been interpreted as containing a broad Lorentzian part (dashed blue curve) and noise. The choice of a Lorentzian function here is arbitrary - almost any function will provide a similar picture, provided that the background has little curvature relative to band features, and is therefore restricted to small inverse momentum in Fourier space. The intensity of this background component has been scaled up by a factor of $A$ to match the background intensity seen beneath the Fermi level, however the portion attributed to noise has subsequently been scaled down by $A^{-1/2}$ as expected for statistical noise. The graphene Dirac band is shown as a line with constant slope in the log-linear plot, representing the Fourier transform of a Lorentzian with the experimentally observed momentum-axis contour (width and amplitude). A similar linear trend is expected for any band with non-zero slope, though non-Lorentzian line shape factors such as resolution broadening will create deviations from a perfect linear trend. 

\subsection{II.B The SDI filter in Fourier space}

Applying a negative second derivative to the data is equivalent to weighting the Fourier components by a factor of $T(\xi)=\xi^2$.  That is to say, given the Fourier decomposition:

\begin{equation}
f(x)=\sum_\xi \hat{f}(\xi)e^{i\xi x},
\end{equation}

the SDI function is simply

\begin{equation}
-f(x)''=\sum_\xi T(\xi) \hat{f}(\xi)e^{i\xi x}.
\label{Eq:filter}
\end{equation}

This weighting factor $T(\xi)$ is illustrated by the dashed green curve in Fig. 1(b). It is easy to see why filtering with $T(\xi)=\xi^2$ is appealing: it has the obvious advantages of suppressing the low frequency background, and sharpening spectral features by giving extra weight to their high frequency components. However, weighting the high frequency noise-dominated region in this way tends to result in completely unusable spectra, as will be shown below in Fig. 2(b,f).

In practice, it is therefore more common to perform a discrete SDI.  If the `dx' infinitesimals in the derivative operator are discretized on a lattice with characteristic spacing $h$, the negative second derivative can be described as:

\begin{equation}
-f(x)''=\frac{2f(x)-f(x+h)-f(x-h)}{h^2}
\end{equation}

Considering this in Fourier space, we have:

\begin{align}
-f(x)'' &=\frac{1}{h^2}\sum_\xi \hat{f}(\xi)[2e^{i\xi x}-e^{i\xi (x+h)}-e^{i\xi (x-h)}] \\
  &= \frac{1}{h^2} \sum_\xi 2(1-cos(h \xi)) \hat{f}(k) e^{i\xi x},
\end{align}

which has the same form as Eq. \ref{Eq:filter}, but with a new weighting factor $T(\xi)=1-cos(h\xi)$. This filter does not explode at high frequencies, and allows high quality images to be achieved by choosing an appropriate value of $h$, such that the first peak of the function overlaps with the good statistical region of the data (see magenta curve in Fig. 1(b)). However, the periodic nature of the $1-cos(h\xi)$ function results in peaks at higher frequencies, which can be expected to contribute noise to the SDI spectrum.

\subsection{II.C Trivial refinements to the SDI filter}

In short, the discrete SDI has the advantage of giving a straightforward mechanism to selectively amplify the `good' frequency band by tuning the discrete spacing parameter $h$, but retains more high frequency noise than is strictly necessary.  Two natural extensions suggest themselves:

\textbf{1. Cutting off high frequency data elements.} We recommend achieving this by setting $T(\xi)=0$ for $h\xi>2\pi$.  Alternatively, it is common to apply a secondary low pass filter (smoothing) to the data, however this has the potential drawback of adding a layer of complexity to the Fourier-space picture.

\textbf{2. Choosing a new underlying function.} The $T(\xi)=1-cos(h\xi)$ function has the disadvantage that the peak width and maximum are defined by the same variable $h$. We recommend defining a Gaussian-like filter instead, so that the peak position and width are associated with independent variables. In this case, one must symmetrize the function about zero, such as by writing:

\begin{equation}
\label{Eq:symGaussian}
T(\xi)=G(\xi-c,\sigma)+G(\xi+c,\sigma)
\end{equation}

Here $G(\xi-c,\sigma)$ is the band pass function (a Gaussian, here) with centroid $c$ and width parameter $\sigma$. For this method, regardless of the function selected, we recommend massaging Eq. \ref{Eq:symGaussian} to set the slope and amplitude to zero at $\xi \rightarrow 0$. For Fig. 1(d), this is achieved as follows: 

\begin{align}
\label{Eq:corrGaussian}
&T(\xi)=-2G(c,\sigma) + \sum_{n=-\infty}^\infty G(\xi-(1+2n)c,\sigma),  \hspace{0.5cm}  |\xi|<c \nonumber  \\ 
&T(\xi)=0                          \hspace{1cm}                                          |\xi|>c
\end{align}

The resulting function contains only two peaks and tails parabolically to zero at $\xi=0,\pm2c$.

\subsection{II.D Applying 1D band pass filters to experimental data}

The SDI-related filters presented above are applied to real ARPES data sets in Fig. 2. The two data sets included here are a background-dominated measurement of the graphene Dirac cone on hBN (Fig. 2(a)), and a multi-band system without obvious background (IrTe$_2$, Fig. 2(e)). The data set in Fig. 2(a) in particular represents an extremely challenging measurement scenario, in which the sample is heavily disordered (as described in Section I) and can only be measured via spectromicroscopy with a $\lesssim$20um beam spot, which is achieved at the cost of some photon flux. Moreover, the single-layer graphene sample is significantly thinner than the photoelectron escape depth, yielding worse signal to noise than one might have with a thicker sample.

The strong background renders the graphene Dirac cone nearly invisible in raw ARPES data (Fig. 2(a)), and the situation is not improved by applying an analytic SDI ($T(\xi)=\xi^2$, see Fig. 2(b)), which shows only noise and vertical stripe artifacts associated with detector imperfections. By contrast, the discrete second derivative approach (Fig. 2(c)) shows the Dirac band quite clearly. This image retains some graininess and vertical stripe artifacts, but is perfectly adequate if one's goal is to trace the band slope. Using a Gaussian filter instead (Fig. 2(d)) improves on this by reducing grain and eliminating some of the narrower stripes, such as show up around $k\sim1\AA$ in Fig. 2(c).

In the IrTe$_2$ spectrum, the bands are easy to see, but they overlap in intensity, creating a visually complex picture. The discrete SDI filter provides a simpler image (Fig. 2(g)), in which it is easy to see that there are 3 bands at the Fermi level. Eliminating the higher frequency periodic replicas of the discrete SDI results in a less grainy image that is otherwise qualitatively identical (Fig. 2(h)).

In spite of successfully sharpening data and dramatically improving signal-to-background by removing low-frequency Fourier components, all of these 1D filters have a significant downside in that the optimal frequency pass window for some band features will not work as well for others. This is particularly clear in the graphene spectrum, where it becomes challenging to see the bands after they flatten out at momenta more than $|k|\gtrsim 0.3\AA$ from the Dirac point. When the band slope becomes shallower, the high-quality region of Fourier space shifts to smaller frequencies that are not passed by the filter. As we will show in the next section, a higher dimensional filter can be used to remove this sensitivity to band slope.

\begin{figure}[t]
\includegraphics[width = 8.7cm]{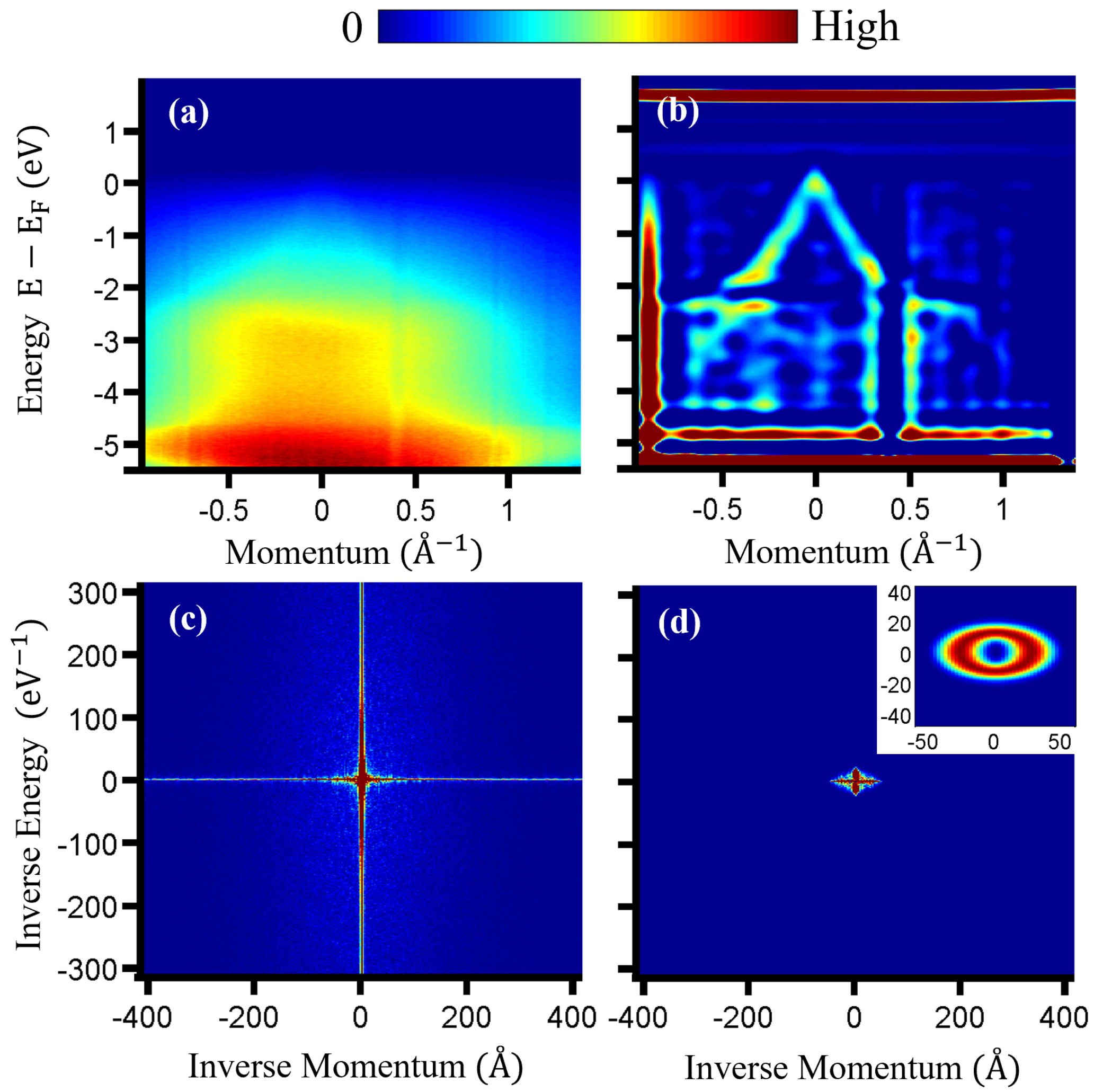}
\caption{{\bf{2D Fourier space of graphene}}: (a) The graphene ARPES data are reproduced from Fig. \ref{fig:1Dfilters}(a). (b) The processed image obtained with a symmetrized 2D Gaussian filter, termed $T$ in the text. (c) The amplitude distribution of a 2D Fourier transform of the data. (d) The amplitude distribution after application of the filter. The amplitude of the filter $T(\xi_k,\xi_E)$ in 2D Fourier space is shown as an inset. The parameters of this filter for graphene are described in the text.}
\label{fig:2Dgraphene}
\end{figure} 

\section{III. Data and band pass filters in higher-dimensional Fourier space}

\subsection{III.A Directly extending the 1D approach}

Modern ARPES data sets are almost always multidimensional, and sometimes incorporate four or more orthogonal dimensions. In this paper we will focus on 2D images with just a single momentum axis and a single energy axis, both for the sake of simplicity, and because this is the most common representation of ARPES measurements. The intuition from the previous section carries over directly to a Fourier transform of 2D (or higher dimensional) data: amplitude in the center of Fourier space tends to be dominated by broad background features, and one can expect amplitude outside of a certain radius to come mostly from statistical noise.

However, there are also ways in which the 2D case \emph{is not} a trivial extension from 1D. A 2D Fourier transform of the background-dominated graphene image (reproduced in Fig. 3(a)) reveals a striking cross-like feature with strong intensity running along the principal axes of 2D Fourier space (Fig 3(c)). This intensity along the axis comes from step functions that effectively occur at the boundaries of the image, because the Fourier transform implicitly (and incorrectly) assumes repeated boundary conditions. Filters that partially crop this cross-like Fourier-space feature will tend to introduce artifacts at the image boundary. Boundary-associated discontinuities have a $1/\xi$-like 1D Fourier decomposition that gives high intensity along the principal axes in a higher-dimensional Fourier space. ($\frac{1}{\xi_E\xi_k}$-like in 2D)

The natural extension of the 1D band pass filters discussed in Section II is a circular band pass filter centered on the origin of the 2D Fourier space (see Fig. 3(d) inset). Distinct parameters are defined for the Gaussian function along the inverse energy ($\xi_E$-axis) and inverse momentum ($\xi_k$-axis) axes. There are many ways to symmetrize the function, and a simple interpolation approach is adopted in this paper: the momentum axes were rescaled into dimensionless units by dividing out the Gaussian centroids, and intensity between the principal axes was then linearly interpolated along a circular contour.  The resulting matrix is termed $T(\xi_k,\xi_E)$, and is represented in the original axis units for convenience.

As identified in Fig. 1, the high signal-to-noise Fourier region is centered around $p_k\sim30$ $\AA$ on the inverse momentum axis. A similar region on the inverse energy axis is found near $p_E=15$ $eV^{-1}$. These values can be expected to resemble the mean free path and scattering lifetime of the imaged band, respectively. The Gaussians were correspondingly centered on 26 $\AA$ and 15 eV$^{-1}$.  

The resulting 2D-filtered plot for graphene is significantly less grainy than 1D filtered images. Unlike the 1D filter, the 2D filter does not strongly link band slope with SDI intensity.  This results in smoother intensity within the Dirac band, which can be more easily followed by eye out to a large momentum of $k\sim0.8\AA^{-1}$. However, strong artifacts are visible along the borders of the image and obscure a much greater fraction of the image than would be the case with a 1D filter.

A similar filter is applied to the IrTe$_2$ data set in Fig. 4. In this case the background is weak, and the cross-like Fourier space feature largely vanishes (see Fig. 4(c)). The high intensity border artifacts seen in the graphene data set (Fig. 3(b)) derive from the interplay of the filter with this cross-like Fourier background feature, and are therefore not present along the SDI panel boundary in Fig. 4(b). Overall, the 2D-filtered image in Fig. 4(b) appears to merely be a less grainy version of the 1D-filtered image in Fig. 2(h).

\subsection{III.B Improving on 2D Fourier filters}

\begin{figure}[t]
\includegraphics[width = 8.7cm]{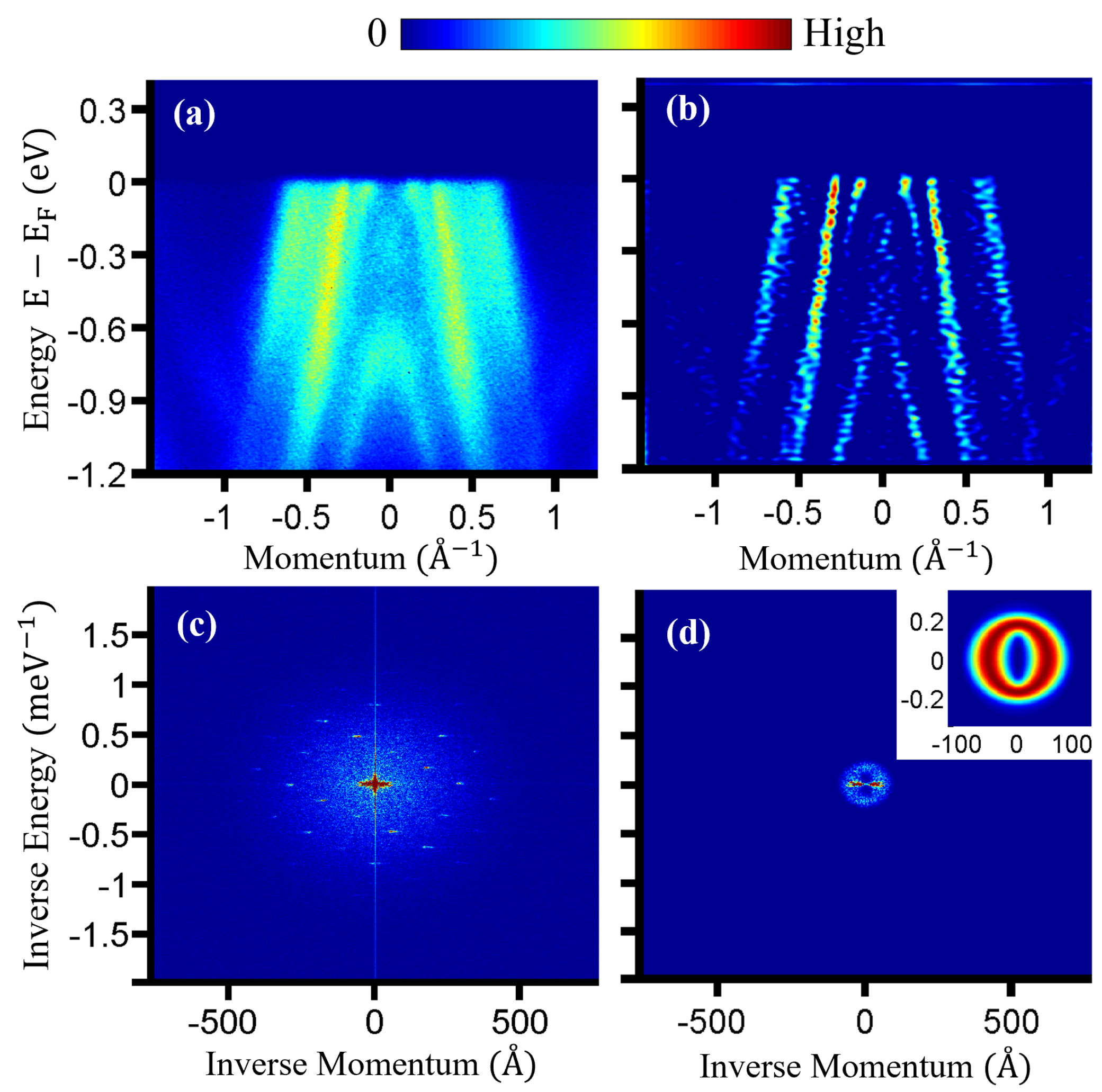}
\caption{{\bf{2D Fourier space of IrTe$_2$}}: (a) The IrTe$_2$ ARPES data are reproduced from Fig. \ref{fig:1Dfilters}(e). (b) The processed image obtained with a symmetrized 2D Gaussian filter, termed $T$ in the text. (c) The amplitude distribution of a 2D Fourier transform of the data. (d) The amplitude distribution after application of the filter. The amplitude distribution of the filter $T$ in 2D Fourier space is shown as an inset. The parameters of this filter for IrTe$_2$ are described in the text.}
\label{fig:2DIrTe2}
\end{figure} 

%***now we can trace the innermost band farther - it is even (weakly) visible after its intersection with the second band from the center
%$H_B=J\times \sigma_z$

\begin{figure*}[t]
\includegraphics[width = 17.5cm]{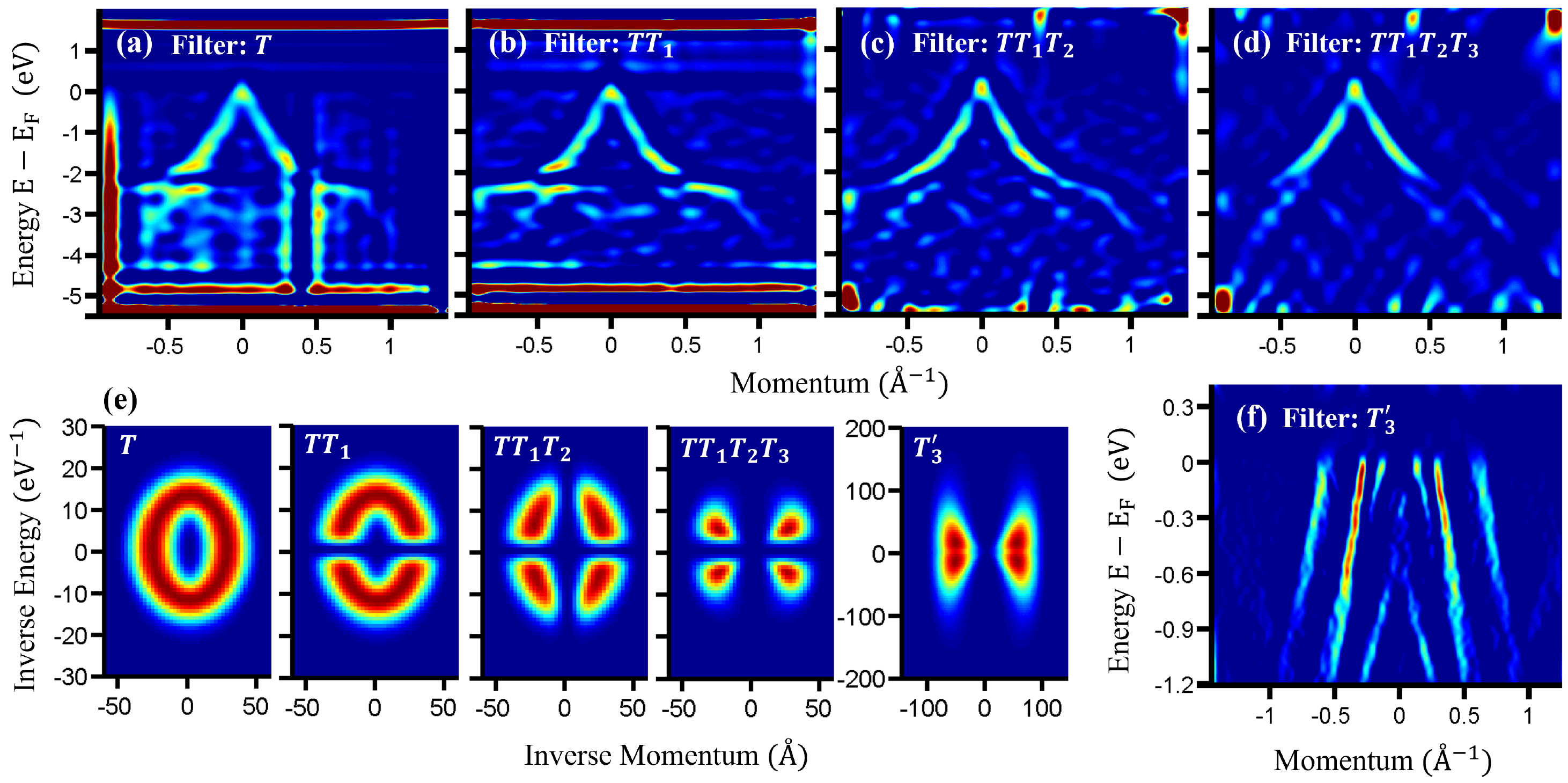}
\caption{{\bf{Background suppressing and feature-selecting improvements}}: (a-d) The positive real part of processed images of graphene with 2D filters $T$, $T \circ T_1$, $T \circ T_1 \circ T_2$ and $T \circ T_1 \circ  T_2 \circ  T_3$, respectively. Here, the overlay of filters is obtained via an entrywise Hademard product [$(A \circ B)_{i,j}=(A)_{i,j}(B)_{i,j}$]. (e) The Fourier space profile of these filters. The rightmost filter is labeled ${T'_3}$, and is applied to enhance InTe$_2$ data in panel (f).}
\label{fig:improvedFilters}
\end{figure*} 

While the above approach has the advantage of being an intuitive extension of the 1D SDI, it does little to make use of the new degree(s) of freedom introduced in higher dimensional data. Several extensions to the 2D filter are explored in Fig. 5, and convey advantages as far as weakening the border artifacts and selectively enhancing features of interest. 

The high-symmetry filter ($T$) described in the previous section is shown at the far left in Fig. 5(e), and the filtered Dirac cone data are reproduced in Fig. 5(a). Given that the boundary artifacts are associated with the cropping of intensity along the principal axes of Fourier-space a natural extension is to remove this intensity entirely. A new filter $T_1$ is defined as follows to remove intensity on the inverse momentum axis:

\begin{equation}
T_1=1-exp \bigg (-\frac{\xi_E^2}{2\sigma_E^2} \bigg )
\end{equation}

Applying this filter on top of the original filter $T$ (as $T \circ T_1$ - see Fig. 5(b)) dramatically improves the representation of the data, as it cleanly removes both the artifact on the left-hand border of the image, and many of the vertical stripe artifacts. The filter is applied with $\sigma_E = 2.52$ $eV^{-1}$ (3 pixels), which was sufficiently large to cleanly remove the border artifacts. It is important to note that the $T_1$ filter will remove \emph{any} completely vertical element of data. While vertical dispersions are usually considered non-physical, they can be a feature of dispersion kinks and `waterfall' spectral features \cite{WF1,WF2,WF3,WF4,WF5,WF6,WF7,WF8,WF9,WF10,WF11} associated with coupling to bosonic modes (magnons, phonons, etc.) and the crossover from a quasiparticle regime near the Fermi level to broader valence bands at higher binding energy.

A complementary filter removing Fourier components along the inverse energy axis is defined as $T_2$, and applied in Fig. 5(c) (using $\sigma_k=10.7$ $\AA$, corresponding to 4 pixels). This filter successfully removes most of the undesired intensity at the upper and lower borders of the image, but has the detrimental effect of suppressing dispersionless (flat) spectral features. As a consequence, the flat dispersion around $E\sim 2.5$ eV binding energy largely vanishes. A final filter $T_3$ selects specifically for the Dirac band, and results in a spectrum in which all other features (artifact or otherwise) have been mostly removed. This filter is defined as:

\begin{equation}
T_3^{(\theta_0)}=exp \bigg \{ -\frac{[\theta-\theta_0-\pi H(\theta-\pi)]^2}{2\sigma_\theta^2} \bigg \} (0 \leq \theta \leq 2\pi)
\end{equation}

Here, $\theta_0$ is a polar angle in unitless Fourier space. Each Fourier axis is rescaled to neutral units by multiplying in the size of the relevant real-space dimension ($\xi'=L\xi / 2\pi$, yielding a discrete spacing of $\Delta \xi'=1$). The Heaviside step function is represented as H(x). Features with constant slope can be highlighted by selecting an angle $\theta_0$ such that Fourier components with this polar angle describe phase and amplitude modulations normal to the band dispersion. In Fig. 5, the $T_3$ filter is further symmetrized across the inverse energy axis, to allow both the left and right branches of the Dirac cone to be observed. The filter parameters were $\theta_0=0.18\pi$ and $\sigma=0.13\pi$ for graphene in Fig. 5(d). A similar $T_3$ filter highlighting Fourier components close to the inverse momentum axis ($\theta_0=0.05\pi$, $\sigma=0.15\pi$) can be used to observe all of the steeply dispersing bands in the InTe$_2$ data set (Fig. 5(f)), and in fact provides a cleaner image than the earlier filters applied to this data set.

It is important to bear in mind that the $T_3$ filter highlights features with a specific slope, and runs an enhanced risk of eliminating or biasing subtle structures in the ARPES spectrum. When using $T_3$ to improve image quality, one should refrain from performing detailed band structure analysis on the filtered image.

\section{IV. Summary}

We have shown that by visualizing the widely applied SDI analysis approach as the application of a band pass filter in Fourier space, one gains a very intuitive picture of factors contributing to image quality. Background features tend to lie in the low frequency Fourier region, and in on-axis `cross' features of higher dimensional Fourier transforms. Statistical noise contributes a high frequency floor to the Fourier decomposition. High quality images are obtained by centering the filter on intermediate frequencies. In particular, good results tend to be achieved with a band pass near the mean free path (inverse momentum axis) or scattering lifetime (inverse energy axis) of a band one wishes to trace.

We find that improved results can be obtained by cropping higher harmonics of the SDI filter, or by using a different form for the band pass. A Gaussian function is suggested, as a way to decouple the centroid and width parameters of the filter. Natural extensions of the SDI approach to 2D (and higher) data sets are discussed, and found to lead to benefits both for image clarity, and for the visibility of band structures with variable band slopes. 

\textbf{Acknowledgements:} This research used resources of the Center for Functional Nanomaterials and the National Synchrotron Light Source II, which are U.S. DOE Office of Science facilities at Brookhaven National Laboratory, under Contract No. DE-SC0012704. The Advanced Light Source is supported by the Director, Office of Science, Office of Basic Energy Sciences, of the U.S. Department of Energy under Contract No. DE-AC02-05CH11231. Work at NYU was supported by the MRSEC Program of the National Science Foundation under Award Number DMR-1420073. Growth of hexagonal boron nitride crystals by K.W. and T.T. was supported by the Elemental Strategy Initiative conducted by the MEXT, Japan and the CREST (JPMJCR15F3), JST.

%XSM: justify "each defect binds two electrons"
%XSM: justify just 1% f-symmetry mixing
%SM: justify the density of vacancy defects (Intrinsic carrier densities for stoichiometrically grown samples in this material family are typically on the order of $1\times 10^{19}$ $cm^-3$, and are largely attributed to point lattice defects such as Se vacancies, ***(2e19/2)/((480.580e-24/3)^-1) ***and correspond to a defect density of $\rho\lesssim0.1$ per formula unit in the outermost quintuple layer if the lattice)
%SM: need spin texture for $\rh0=0.18\%$

% (1) 1$\%$ f-wave admixture;  (2) typical defect density in crystals; (3) each defect binds 2 electrons into the resonance state (4) how participation ratio/LDOS is obtained; (4) different cluster sizes do not change the result; (5) difference between $K_r$ and $K_x$

\end{document}